\documentclass[aps,pre,reprint,superscriptaddress,amsmath,amssymb,showpacs,floatfix,longbibliography]{revtex4-1}
\usepackage{graphicx}
\usepackage{bm}
\usepackage{color}
\usepackage{lipsum}

\begin{document}

\title{Effect of a magnetic field on the thermodynamic uncertainty relation}

\author{Hyun-Myung Chun}
\affiliation{II. Institut f{\"u}r Theoretische Physik, Universit{\"a}t Stuttgart,
70550 Stuttgart, Germany}
\author{Lukas P. Fischer}
\affiliation{II. Institut f{\"u}r Theoretische Physik, Universit{\"a}t Stuttgart,
70550 Stuttgart, Germany}
\author{Udo Seifert}
\affiliation{II. Institut f{\"u}r Theoretische Physik, Universit{\"a}t Stuttgart,
70550 Stuttgart, Germany}

\date{\today}

\begin{abstract}
The thermodynamic uncertainty relation provides a universal lower bound
on the product of entropy production and the fluctuations of any
current. While proven for Markov dynamics on a discrete set of states
and for overdamped Langevin dynamics, its status for underdamped
dynamics is still open. We consider a two-dimensional harmonically
confined charged particle in a magnetic field under the action of an
external torque. We show analytically that, depending on the sign of the
magnetic field, the thermodynamic uncertainty relation does not hold 
for the currents associated
with work and heat. A strong magnetic field can effectively localize the
particle with concomitant bounded fluctuations and low dissipation. Numerical
results for a three-dimensional variant and for further currents suggest
that the existence of such a bound depends crucially on the specific
current. 
\end{abstract}
\pacs{}
\maketitle

%%%%%%%%%%%%%%%%%%%%%%%%%%%%%%%%%%%%%%%%%%%%%%%%%%%%%%%%%%%%%%%%%%%%%%%%%%%%%

\section{Introduction}
Nonequilibrium systems are characterized by constantly flowing currents and
inevitably accompanied dissipation.
Some currents, such as the particle current in a molecular motor or the
energy current in a heat engine, can be used for practical purposes.
In order to optimize the practical use of the currents, it is necessary to
suppress two factors: uncertainty and dissipation.
Uncertainty of currents due to thermal fluctuation makes a system less 
predictable and dissipation should be minimal to sustain the efficient 
operation of the system.

Recent studies have revealed that a universal trade-off between 
uncertainty and dissipation exists in nonequilibrium 
systems~\cite{Barato:2015kq,Gingrich:2016ip,Seifert:2018st}.
This trade-off, called the thermodynamic uncertainty relation,
states that suppressing uncertainty costs more dissipation, and conversely, 
reducing dissipation causes larger uncertainty.
Since its first discovery~\cite{Barato:2015kq},
the thermodynamic uncertainty relation has been proven for Markov jump 
dynamics
on a discrete set of states~\cite{Gingrich:2016ip,Proesmans:2017ic,Dechant:2018ea} and Langevin
dynamics~\cite{Polettini:2016hu,Gingrich:2017jm,Dechant:2018ga,
Dechant:2018vu,Dechant:2018ea} for continuous variables.
Generalizations to finite time statistics~\cite{Pietzonka:2017fi,Horowitz:2017ut,Pigolotti:2017ge} and to periodically driven systems~\cite{Proesmans:2017ic,Barato:2018hb,Koyuk:2019kr} have also been 
reported.

The standard proofs of the thermodynamic uncertainty relation, however, assume that
the observables referring to the state of the system are even under time reversal.
The proven validity of the thermodynamic uncertainty relation is thus limited 
to the overdamped regime where inertia is negligible.
The thermodynamic uncertainty relation for systems with variables that
are odd under time reversal, like in underdamped Langevin dynamics, remains a 
puzzling problem.
Two examples have shown intriguing features.
First, for a driven underdamped particle in a periodic potential, no
violation of the thermodynamic uncertainty relation has been observed
for the particle current~\cite{Fischer:2018fi}.
Second, for ballistic transport in multiterminal conductors,
a violation is observed only in the presence of a magnetic
field~\cite{Brandner:2018cr,Macieszczak:2018jv}.
For near-equilibrium systems with broken Onsager symmetry, which can be
caused by magnetic fields, a mathematical explanation for the violation 
was provided in Ref.~\cite{Macieszczak:2018jv}.
In addition, a weaker trade-off for underdamped Langevin dynamics,
in which dissipation is replaced by a quantity called dynamical 
activity~\cite{Fischer:2018fi, VanVu:2019uu} has been reported.
% and a lower bound on uncertainty based on the 
% fluctuation theorem have been derived recently~\cite{Hasegawa:2019to}.

In this paper, we investigate a physical mechanism by which a magnetic field
acting on an underdamped particle
causes a violation of the thermodynamic uncertainty relation.
To this end, we consider a charged Brownian particle driven by
an external torque in the presence of a magnetic field~\cite{Lee:2019wl}. 
We focus on the work done by the external torque on the particle 
which can be written in form of an integrated current. 
We show that the Lorentz force induced by a strong magnetic field 
in underdamped dynamics localizes the motion of the particle into a 
smaller area, which in this case leads to a decrease of dissipation.
On the other hand, the uncertainty of the work current does not increase
with a strong magnetic field, due to a fast decrease of current fluctuations.
Thus a strong magnetic field can decrease the dissipation without increasing 
the uncertainty, which results in a violation of the thermodynamic uncertainty 
relation.

\section{Model}

We first consider two-dimensional motion of a charged particle 
in a fluid in equilibrium at temperature $T$.
The position and the velocity of the particle are denoted by column vectors 
$\bm{x} = (x_1, x_2)^{\rm T}$ and $\bm{v} = (v_1,v_2)^{\rm T}$ where the 
superscript `${\rm T}$' stands for the transpose.
The particle is trapped by a conservative force associated with a harmonic 
potential $V(\bm{x}) = k\bm{x}^2/2$ with a spring constant $k$.
A constant magnetic field $\bm{B}$ of strength $B=|\bm{B}|$, 
whose direction is perpendicular to the plane of the particle motion,
induces a magnetic Lorentz force $\bm{g}(\bm{v}) = qB(v_2, -v_1)^{\rm T}$
with the charge of the particle $q$.
In addition to these forces, an external torque
$\bm{f}(\bm{x}) = \kappa(x_2, -x_1)^{\rm T}$ now drives the particle out of 
equilibrium, resulting in a circular particle current. 
The overdamped analogon of this system without a magnetic field
has been thoroughly studied~\cite{Filliger:2007vr,Kwon:2011th,Pietzonka:2018de} 
and experimentally realized~\cite{Chiang:2017fy,Argun:2017hb}.

Overall, the motion of the particle is governed by the underdamped Langevin equation
\begin{equation}
\begin{aligned}
    \dot{\bm{x}}(t) & = \bm{v}(t), \\
    m \dot{\bm{v}}(t) & = -k\bm{x}(t) + \bm{f}(\bm{x}(t)) - \gamma \bm{v}(t)
    + \bm{g}(\bm{v}(t)) + \bm{\xi}(t),
\end{aligned}
\end{equation}
where $m$ is the mass of the particle and $\gamma$ is a damping coefficient.
The Gaussian white noise $\bm{\xi}(t) = (\xi_1(t),\xi_2(t))^{\rm T}$ is 
characterized by $\langle \xi_i(t) \rangle = 0$ and
$\langle \xi_i(t) \xi_j(t') \rangle = 2\gamma k_\mathrm{B} T \delta_{ij}\delta(t-t')$
with the Boltzmann constant $k_\mathrm{B}$.
The angle brackets $\langle \cdot \rangle$ stand for the ensemble average.
Hereafter, we set the charge of the particle $q$ and the Boltzmann constant $k_\mathrm{B}$
to unity.
For brevity, we introduce collective notations
$\bm{z} = (\bm{x},\bm{v})^{\rm T}$ and $\bm{\eta} = (\bm{0},\bm{\xi})^{\rm T}$
so that the equations of motion are concisely written as
\begin{equation}\label{eq:Langevin}
    \dot{\bm{z}}(t) = -\mathsf{A}\bm{z}(t) + \bm{\eta}(t)
\end{equation}
with
\begin{equation}\label{eq:matA}
    \mathsf{A} = \frac{1}{m} \begin{pmatrix}
    0 & 0 & -m & 0 \\
    0 & 0 & 0 & -m \\
    k & -\kappa & \gamma & -B \\
    \kappa & k & B & \gamma
    \end{pmatrix}.
\end{equation}
The Gaussian white noise $\bm{\eta}(t)$ is characterized by
$\langle \eta_i(t) \rangle = 0$ and
$\langle \eta_i(t) \eta_j(t') \rangle = 2D_{ij}\delta(t-t')$
where $D_{ij}$ are the elements of the matrix
$\mathsf{D} = (\gamma T/m^2) \rm diag\{ 0,0,1,1 \}$.
This linear Langevin equation is a multivariate 
Ornstein-Uhlenbeck process~\cite{Gardiner:2010tp}. 

When the real parts of all eigenvalues of the matrix $\mathsf{A}$ are 
positive, the system is stable and the probability distribution of $\bm{z}$ 
eventually converges to a steady state. This stability condition of the system is given by
(see Appendix~\ref{sec:stable_condition} for details)
\begin{equation}\label{eq:stability}
    \gamma k + \kappa B - \kappa^2 m/\gamma > 0.
\end{equation}
Consequently, the particle is confined by the potential if the strength of torque 
$|\kappa|$ is moderate enough to satisfy the stability condition,
otherwise the particle escapes from the potential.
%Since the linear equation makes the model analytically tractable,
%similar models have been used to describe some physical 
%phenomena~\cite{Chun:2015ua,Park:2016ig,Chun:2018de}.
%Overdamped systems involving only position variables have been thoroughly
%studied in~\cite{Kwon:2011th} at the framework of stochastic thermodynamics.
%One interesting application of the overdamped Brownian motion, 
%so-called Brownian gyrator, has been proposed~\cite{ger:2007vr} and
%experimentally realized~\cite{Chiang:2017fy,Argun:2017hb}.

%%%%%%%%%%%%%%%%%%%%%%%%%%%%%%%%%%%%%%%%%%%%%%%%%%%%%%%%%%%%%%%%%%%%%%%%%%%%%
\begin{figure}
\includegraphics*[width=\columnwidth]{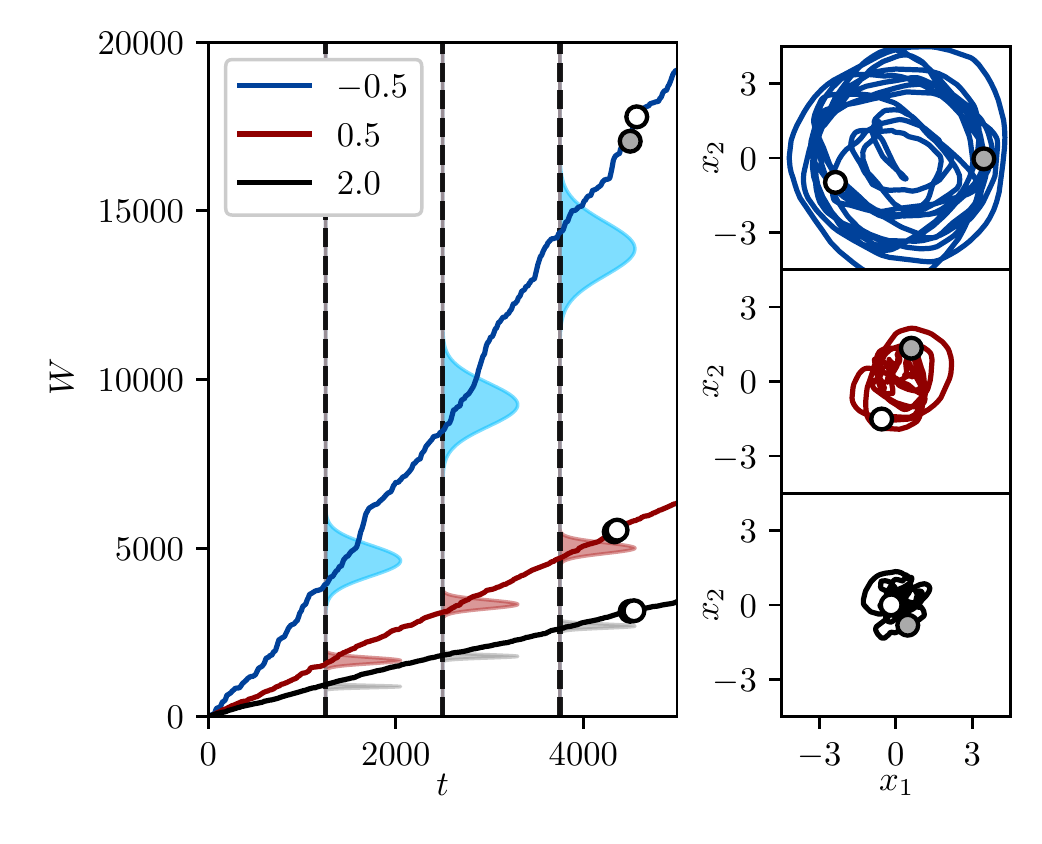}
\caption{
The left panel shows three sample trajectories of the work done by an external
torque on a charged Brownian particle for different strengths of the 
magnetic field $B$.
The strengths of the magnetic field are $-0.5, 0.5$, and $2.0$ from top to 
bottom in both panels.
The distributions behind the time series represent the distributions of
the work done up to the given time.
The other parameters are fixed as $m = 1$, $\gamma = 1$, $k = 2$,
$\kappa = 1$, and $T = 1$.
The right panel shows typical trajectories of the particle corresponding to 
the three different strengths of the magnetic field.
The gray (white) circle denotes the start (end) point of each 
trajectory.
}
\label{fig:trajectories}
\end{figure}
%%%%%%%%%%%%%%%%%%%%%%%%%%%%%%%%%%%%%%%%%%%%%%%%%%%%%%%%%%%%%%%%%%%%%%%%%%%%%

The external torque does work on the particle with rate
\begin{equation}\label{eq:workrate}
	\dot{W}(t) = \bm{f}(t) \cdot \bm{v}(t)
= \kappa(x_2(t) v_1(t) - x_1(t) v_2(t))	
\end{equation}
This work either increases the energy of the particle or is 
dissipated into the surrounding as heat with rate
$\dot{Q}(t) = (\gamma \bm{v}(t) - \bm{\xi}(t)) \circ
\bm{v}(t)$~\cite{Sekimoto:1998uf,Seifert:2012es}
where the symbol $\circ$ denotes the Stratonovich product.
With the total energy of the particle
$E = m \bm{v}^2/2 + k \bm{x}^2/2$,
these definitions are consistent with the first law of thermodynamics 
$\dot{E}(t) = \dot{W}(t) - \dot{Q}(t)$.

The mean dissipation in a nonequilibrium system is quantified by
the entropy production rate $\sigma$.
The total entropy production consists of the change in the stochastic entropy 
of the system and the medium entropy production which is given by
$S_{\rm med} = Q(t) /T$ in our case~\cite{Seifert:2012es}.
In a steady state, both the stochastic entropy and the total energy of the 
system do on average not change over time.
Thus the mean entropy production rate is given by $\sigma
= \langle \dot{S}_{\rm med} \rangle_{\rm ss}
= \langle \dot{Q} \rangle_{\rm ss} / T
= \langle \dot{W} \rangle_{\rm ss}/T $.
The subscript of $\langle \cdot \rangle_{\rm ss}$ indicates that the ensemble
average is taken in a steady state which we will also take as initial distribution.

In the following section, we analyze the behavior of the rate of work as a 
current in order to investigate the effect of a magnetic field on the 
thermodynamic uncertainty relation.
Since the work current is proportional to the angular momentum of
the particle, it can be interpreted as a measure of the circular particle current.
Also, the work current is equivalent to the heat current as the mean value
and diffusion coefficient of the heat current coincides with 
those of the work current in the steady state.

\section{Effect of a magnetic field on the thermodynamic uncertainty relation}

The integrated work current $W(t) = \int_0^t dt' \dot{W}(t')$ depends on the
trajectory of the particle and thus is
a stochastic variable. As a consequence of the
torque, the work $W(t)$ increases on average as also visible in
Fig.~\ref{fig:trajectories}.
The deviation of the work from its mean value is quantified through the 
diffusion coefficient
\begin{equation}\label{eq:Dw_def}
    D_{W} = \lim_{t\to\infty}
    \frac{1}{2t}\left(
    \langle W(t)^2 \rangle_{\rm ss} - \langle W(t) \rangle^2_{\rm ss}
    \right).
\end{equation}
The uncertainty of the fluctuating work current is characterized by the 
squared relative uncertainty
\begin{equation}\label{eq:rel_unc}
	\epsilon_W^2 = 2 D_{W} / \langle \dot{W} \rangle_{\rm ss}^2.
\end{equation}
It measures how fast the variance increases with respect to the mean growth
rate of the integrated current.
The thermodynamic uncertainty relation states that the quantity
$\mathcal{Q}_W = \epsilon_W^2 \sigma/2$, which we shall refer to
as the uncertainty product, is bounded from below by the Boltzmann constant 
which is set to unity here. 

In polar coordinates, the work current is given by
$\dot{W} = -\kappa r^2 \dot{\theta}$ with the radial distance $r = |\bm{x}|$
and the polar angle $\theta = \tan^{-1}(x_2/x_1)$.
The steady state probability distribution of the Ornstein-Uhlenbeck process
can be obtained by standard 
procedures~\cite{Gardiner:2010tp,Kwon:2011th,Lee:2019wl}.
For a given $r$, the conditional distribution of the angular velocity 
$\dot{\theta}$ is a Gaussian with mean value $-\kappa/\gamma$ 
and variance $T/(mr^2)$.
Thus the mean work current is proportional to the mean squared
radial distance, which is given by
\begin{equation}\label{eq:r_sq}
    \langle \dot{W} \rangle_{\rm ss}
    = \frac{\kappa^2}{\gamma} \left\langle r^2 \right\rangle_{\rm ss}
    = \frac{2\kappa^2 T}{\gamma k + \kappa B - \kappa^2 m /\gamma}
\end{equation}
as derived in detail in Appendix~\ref{sec:stable_condition}.
The stability condition \eqref{eq:stability} guarantees that 
\eqref{eq:r_sq} is always positive.
Since the mean values of the work and heat current are balanced
in the steady state, the mean entropy production rate scales as
$\sigma = \langle \dot{W} \rangle_{\rm ss}/T \sim B^{-1}$ for large $B$ 
provided that $\kappa B > 0$.

This decrease of dissipation with increasing strength of the magnetic field is
a result of a stronger localized motion of the particle.
In the absence of a magnetic field, the sign of $\kappa$ determines
the rotational direction of the motion.
A positive $\kappa$ leads to a clockwise rotation around the center of 
potential.
By supplying energy to the particle, the external torque only increases
the radial distance.
Once the magnetic field is turned on, the induced Lorentz force pushes the 
particle either inward or outward of the potential depending on the sign
of $\kappa B$ as shown in the right panels of Fig.~\ref{fig:trajectories}.
When $\kappa B < 0$, the magnetic field reinforces the tendency to increase
the radial distance.
When $\kappa B > 0$, however, the Lorentz force makes the particle prefer to 
head towards the center of the potential.
Thus, as $B$ increases, the particle becomes localized in a smaller area 
around the center.
This also reduces the dissipation since 
the work rate, Eq.~\eqref{eq:r_sq}, and thus also the heat rate, decrease with
the particle being localized in a smaller area.
The localization effect only occurs in the presence of both the external 
torque and the magnetic field.

On the other hand, the squared relative uncertainty does not increase with
stronger $B$. The expressions for the mean current and the diffusion
coefficient show that the relative uncertainty even decreases with stronger 
$B$ in the small mass limit because the 
diffusion coefficient decreases much faster than the mean current. 
Since the Langevin equation is linear, the exact expressions for the 
mean current and the 
diffusion coefficient can be obtained by expanding the scaled cumulant generating function~\cite{Pietzonka:2016un,Touchette:2018fb,Pietzonka:2018de}
as shown in detail in Appendix~\ref{sec:exact_expression}. 
In terms of the dimensionless parameters
$\kappa_0 = \kappa/k$, $B_0 = B/\gamma$ and $m_0 = k m/\gamma^2$,
they turn out to be given by
\begin{equation}\label{eq:wss}
    \langle \dot{W} \rangle_{\rm ss}
    = \frac{2 \kappa_0^2}{1 + \kappa_0 B_0 - \kappa_0^2 m_0}
    \left( \frac{T}{\tau} \right),
\end{equation}
and
\begin{equation}\label{eq:Dw}
    D_W
    = \frac{2 \kappa_0^2 [1 + \kappa_0^2 (1 + 3 m_0) + \kappa_0^3 m_0 B_0]}
    {(1 + \kappa_0 B_0 - \kappa_0^2 m_0)^3}
    \left( \frac{T^2}{\tau} \right),
\end{equation}
where $\tau = \gamma/k$ is a parameter with the dimension of time. While the 
leading term in $D_W$ for large magnetic fields is $(2 \kappa_0^2 m_0 T^2) / (\tau B_0^{2})$, 
the mean work rate only scales with $1/B_0$. As a consequence the 
relative uncertainty $\epsilon_W^2$, Eq.~\eqref{eq:rel_unc},
converges to a constant as $B_0$ increases. In the small mass limit $m_0 \rightarrow 0$
the leading order of $D_W$ even becomes $B_0^{-3}$ which leads to a decreasing relative uncertainty
in this limit. An alternative derivation of the scaling order of $D_W$ 
is presented in Appendix~\ref{sec:scaling}.
Overall, the converging or decreasing relative uncertainty 
$\epsilon_W^2$ for large magnetic fields cannot balance the decrease 
of the mean entropy production $\sigma$. As a consequence the
thermodynamic uncertainty relation can be broken in the presence of a 
strong magnetic field.

\begin{figure}
\includegraphics*[width=\columnwidth]{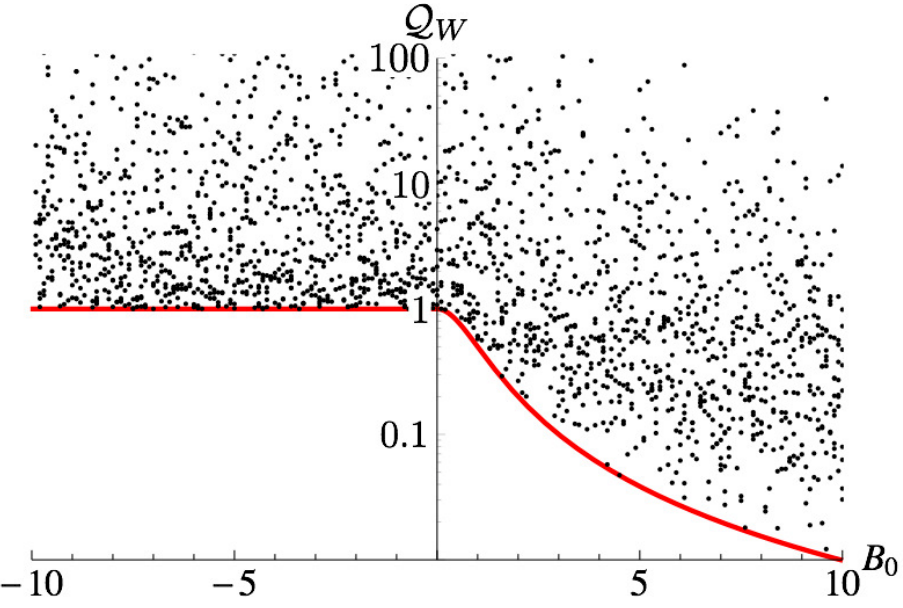}
\caption{
The scatter plot shows that the uncertainty product is always greater than
the bound depicted as solid red line. The bound takes on 1 for
$\kappa_0 B_0 < 0$ 
and otherwise corresponds to Eq.~\eqref{eq:TUR_bound}.
Each data point is determined by a parameter set $(\kappa_0,m_0,B_0)$
that satisfies the stability condition.
The parameters $\kappa_0 \in [0, 10]$ and $m_0 \in [0, 1]$ are randomly 
chosen, while $B_0 \in [-10,10]$ is chosen at a constant interval of 0.1.
}
\label{fig:lower_bound}
\end{figure}

Using the expressions for the mean current and the diffusion coefficient, 
Eqs.~\eqref{eq:wss} and \eqref{eq:Dw},
the uncertainty product can be explicitly written as
\begin{equation}\label{eq:U_product}
    \mathcal{Q}_W
    = \frac{1 + \kappa_0^2 (1 + 3m_0) + \kappa_0^3 m_0 B_0}
    {(1 + \kappa_0 B_0 - \kappa_0^2 m_0)^2}.
\end{equation}
Without the magnetic field ($B_0 = 0$), the thermodynamic uncertainty 
relation $\mathcal{Q}_W \geq 1$ is restored.
For given $\kappa_0$ and $B_0$,
the uncertainty product is monotonously decreasing with decreasing $m_0$.
Thus the uncertainty product is greater than
\begin{equation}\label{eq:small_inertia}
    \lim_{m_0\to0} \mathcal{Q}_W
    = \frac{1+\kappa_0^2}{(1+\kappa_0 B_0)^2},
\end{equation}
which is greater than $1+\kappa_0^2$ for $\kappa_0 B_0 \leq 0$.
Consequently, the underdamped model with $\kappa_0 B_0 \leq 0$ obeys the
usual bound on $\mathcal{Q}_W$ as shown in Fig.~\ref{fig:lower_bound}. 
%The limit $m_0 \to 0$ with a finite $B_0$ is a singular limit 
%that can be described by an effective Langevin equation with a nonwhite 
%noise~\cite{Chun:2018de}.
For $\kappa_0 B_0 > 0$, the small inertia bound in \eqref{eq:small_inertia} is
minimal for $\kappa_0 = B_0$ and then simply reads
\begin{equation}\label{eq:TUR_bound}
    \mathcal{Q}_W \geq \frac{1}{1 + B_0^2},
\end{equation}
where equality holds when $\kappa_0 = B_0 = 0$.
Following the optimizations done so far the bound \eqref{eq:TUR_bound} 
becomes saturated when $\kappa_0 = B_0$ and
$m_0 \rightarrow 0$.
As sketched in the positive half of Fig.~\ref{fig:lower_bound}, the bound 
goes below 1 and approaches zero for large $B_0$.
In this regime the thermodynamic uncertainty relation does not hold and the 
uncertainty product can even approach zero. We have thus shown that the 
localization of the motion of the particle due to the magnetic field, which 
occurs only when $\kappa_0 B_0 > 0$, is responsible for the violation of 
thermodynamic uncertainty relation in \eqref{eq:TUR_bound}.

We briefly discuss further inequalities that have been derived in recent 
works.
A bound obtained from the fluctuation theorem for entropy production for 
general dynamics~\cite{Hasegawa:2019to} can be applied to our model 
only when there is no magnetic field and hence the fluctuation theorem 
holds.
On the other hand, another inequality termed hysteretic thermodynamic 
uncertainty relation~\cite{Proesmans:2019td}, which takes into account
the time-reversed dynamics, can be applied to our model for a weak enough 
magnetic field.
For strong magnetic fields, however, the time-reversed dynamics can
become unstable as a consequence of the asymmetry with respect to $B$
in the stability condition in \eqref{eq:stability}.
In their respective range of validity both inequalities provide 
exponentially weaker lower bounds on the uncertainty product than 
\eqref{eq:TUR_bound} does.

%%%%%%%%%%%%%%%%%%%%%%%%%%%%%%%%%%%%%%%%%%%%%%%%%%%%%%%%%%%%%%%%%%%%%%%%%%%%%
%%%%%%%%%%%%%%%%%%%%%%%%%%%%%%%%%%%%%%%%%%%%%%%%%%%%%%%%%%%%%%%%%%%%%%%%%%%%%

\section{Uncertainty product for other currents}

So far we have focused on the work current.
In general one can consider arbitrary currents.
The lower bound on the uncertainty product then can depend 
on the current of interest.
For example, for ballistic transport in multiterminal conductors,
the uncertainty product of the particle current is bounded from below by
a non-zero constant~\cite{Brandner:2018cr}, whereas that of a generalized 
current defined in the linear response theory can come arbitrarily close
to 0 by optimizing the chemical potentials applied to the 
conductor~\cite{Macieszczak:2018jv}. Based on numerical results
we show that a similar effect can be observed for the present model as well.

\begin{figure}
	\includegraphics[scale=1]{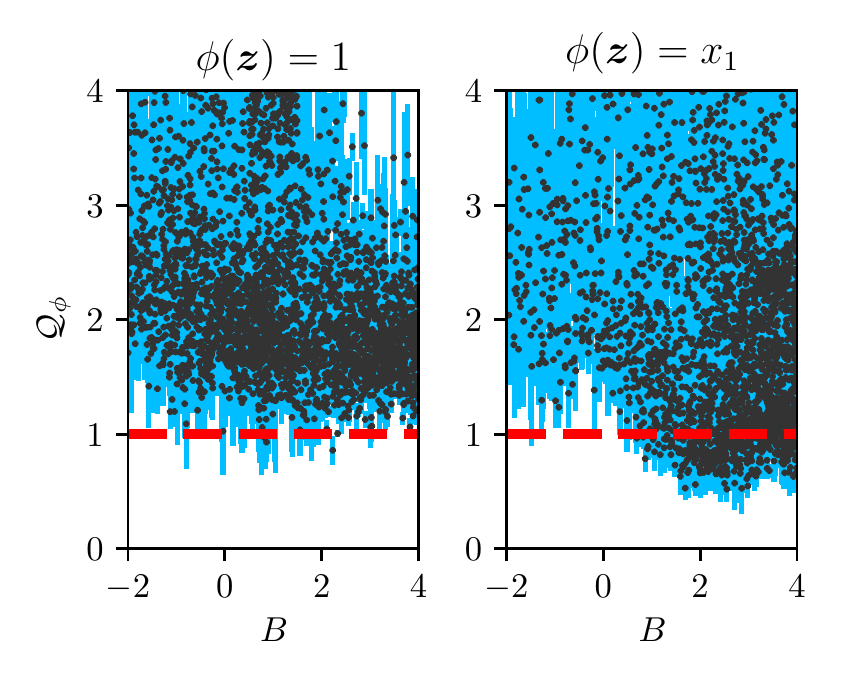}
	\caption{\label{fig:windingCurrents} Numerical results for the uncertainty product $\mathcal{Q}_\phi$ for the winding number current \eqref{eq:windingNumber} with two different weighting factors 
 $\phi(\bm{z}) = 1$ (left) and $\phi(\bm{z}) = x_1$ (right). Each dot represents one stable parameter set with $B \in [-2, 4]$, $\kappa \in [0.05, 5]$, $k \in [0, 4]$, $\gamma \in [0.1, 10]$, $T \in [0.5, 1.5]$. The mass $m$ is set to unity. The vertical blue lines are an estimator for the error.
 The error is estimated by calculating the $25\,\%$ percentile for the slope of the mean current and its variance on an ensemble of trajectories. The entropy production rate $\sigma$ is calculated on base of the analytical results Eq.~\eqref{eq:r_sq}. The error of the uncertainty product then follows by propagation of error from the error of the mean current and the error of the diffusion coefficient. }
\end{figure}

In the case of the charged Brownian particle, the work current
$\dot{W} = -\kappa r^2 \dot{\theta}$ can be interpreted as 
an angular current $\dot{\theta}$ with a weighting factor $-\kappa r^2$.
An alternative way of quantifying the angular current is to look
at the more general class of winding number currents that count the number of 
crossings of the positive $x_1$-axis with a weighting factor $\phi(\bm{z})$.
Formally this current is given by
\begin{equation}
\label{eq:windingNumber}
\begin{aligned}
    j_\phi(t) 
    & = \frac{1}{t}\int_0^t ds ~ \phi(\bm{z}(s))
    \chi_{x_1(s^-)}\chi_{x_1(s^+)} \\
    & ~~~ \times \left[ 
    \chi_{x_2(s^-)}(1 - \chi_{x_2(s^+)})
    - (1 - \chi_{x_2(s^-)}) \chi_{x_2(s^+)}
    \right]
\end{aligned}
\end{equation}
where $\chi_x$ is an indicator that is 1 if $x>0$ and 0 otherwise.
The times $s^-$ and $s^+$ denote the time right before and after $s$, 
respectively.
The advantage of the winding number current is that it could be directly measured in 
experimental situations without tracing complete trajectories of the particle.
The uncertainty product associated with the current $j_\phi$ is given as
\begin{equation}
	\mathcal{Q}_\phi = \frac{ D_\phi }{\langle j_\phi \rangle_\text{ss}^2} \sigma
\end{equation}
where $D_\phi$ is the diffusion coefficient of the time integrated current $J_\phi(t) = \int_0^t dt' j_\phi(t')$.
We numerically calculate the winding number current for two 
different weighting factors $\phi(\bm{z}) = 1$ and $\phi(\bm{z}) = x_1$.
A conceptually similar current as the former one was recently used in~\cite{MarslandIII:2019un} in the context of biochemical oscillations.
The uncertainty products for both weights are depicted in 
Fig.~\ref{fig:windingCurrents} for random parameters. The results show that
the uncertainty product goes well below 1 in the latter case.
In the former case, however, it seems to be bounded from below by 1 in the 
margin of error.
In this special case the contribution to the current is independent from the
radial distance and thus the localization of motion does not affect the current
as strongly.

%%%%%%%%%%%%%%%%%%%%%%%%%%%%%%%%%%%%%%%%%%%%%%%%%%%%%%%%%%%%%%%%%%%%%%%%%%%%%
\begin{figure}
\includegraphics*[width=\columnwidth]{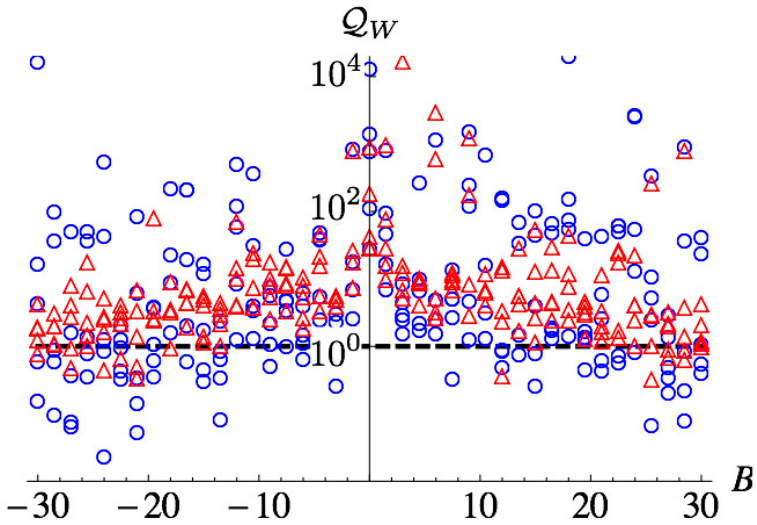}
\caption{
Scatter plots of the uncertainty products for work done by $\bm{f}_{12}$
(blue circle) and $\bm{f}_{23}$ (red triangle) in three-dimensional motion 
with a constant magnetic field $\bm{B}=(0,0,B)$.
Each data point is determined by a parameter set 
$(\kappa_{12},\kappa_{23},\kappa_{31},m,B)$ that satisfies the stability 
condition.
The parameters $\kappa_{12}, \kappa_{23}, \kappa_{31} \in [0,10]$ and
$m \in [0,1]$ are randomly chosen,
while $B \in [-30,30]$ is chosen at a constant interval of 1.5.
The other parameters are fixed as $\gamma = k = T = 1$.
}
\label{fig3}
\end{figure}
%%%%%%%%%%%%%%%%%%%%%%%%%%%%%%%%%%%%%%%%%%%%%%%%%%%%%%%%%%%%%%%%%%%%%%%%%%%%%

Next we consider a three-dimensional variant of the model.
The motion of the particle is then described by the position
$\bm{x} = (x_1, x_2, x_3)^{\rm T}$ and
the velocity $\bm{v} = (v_1, v_2, v_3)^{\rm T}$.
Three mutually perpendicular external torques are applied to the particle.
They are given by $\bm{f}_{12} = \kappa_{12}(x_2,-x_1,0)^{\rm T}$,
$\bm{f}_{23} = \kappa_{23}(0,x_3,-x_2)^{\rm T}$,
and $\bm{f}_{31} = \kappa_{31}(x_3,0,-x_1)^{\rm T}$, respectively.
The strengths of torques $\kappa_{12}$, $\kappa_{23}$, and $\kappa_{31}$
are in general different from each other.
Fig.~\ref{fig3} shows numerically calculated uncertainty products
of the work done by the torques $\bm{f}_{12}$ and $\bm{f}_{23}$ in the 
presence of a constant magnetic field $\bm{B}=(0,0,B)$.
In the absence of the magnetic field, no violation of the thermodynamic 
uncertainty relation is observed in both cases. When the magnetic field is 
turned on, the uncertainty product for the work done by either torque 
goes below the usual bound of 1. This is due to the localization of the 
projected motion onto the $x_1x_2$-plane and $x_2 x_3$-plane.
However, the localization effect on the $x_1x_2$-plane is stronger.
As a consequence, the uncertainty product 
of the work done by $\bm{f}_{12}$ decreases more rapidly
compared to the  work done by $\bm{f}_{23}$.
These numerical case studies suggest that the lower bound on the 
uncertainty product depends on the current of interest.

\section{Concluding perspective}

We have investigated the effect of a magnetic field on the thermodynamic 
uncertainty relation through a driven charged Brownian particle in
a magnetic field. We have focused on the work current done by an external torque.
 The physical mechanism of the violation of
the thermodynamic uncertainty relation due to a magnetic field arises, since,
on the one hand, the motion of the particle becomes localized due to 
the magnetic Lorentz force, which reduces dissipation.
On the other hand, the uncertainty of the work current does not increase
with increasing strength of the magnetic field.
Thus the usual trade-off between dissipation and uncertainty of the 
work current does not apply in the presence of a magnetic field, 
resulting in a violation of the thermodynamic uncertainty relation.
We also found that the lower bound on the uncertainty product depends on the current of 
interest as indicated by numerical results for winding number 
currents in two-dimensions and for work currents in the three-dimensional 
analogue of the harmonically trapped, driven particle.

This study leaves at least two general questions for future research on
currents in underdamped dynamics in a magnetic field. First, can we classify
the conditions under which the thermodynamic uncertainty relation still
holds? In our two-dimensional system, the relative sign between magnetic
field and torque has been crucial. How, if at all, can this finding be
generalized? Second, for the putative other class, does there exist a 
universal weaker bound that is as transparent
as the thermodynamic uncertainty relation?
Finally, understanding the mechanism behind the violation of the 
thermodynamic uncertainty relation in the presence of a magnetic
field might help to understand its validity for underdamped motion without
a magnetic field.
% that, e.g., generalize our
% ? 
% in general, understanding the violation of the  thermodynamic uncertainty
% relation in the presence of the magnetic field might shine light on the question
% whether it holds for underdamped motion without a magnetic field.

\section*{Acknowledgements}
H.-M.\,C. is supported by the the National Research Foundation of Korea (Grant No. 2018R1A6A3A03010776).

%%%%%%%%%%%%%%%%%%%%%%%%%%%%%%%%%%%%%%%%%%%%%%%%%%%%%%%%%%%%%%%%%%%%%%%%%%%%%
\appendix
\section{Steady state distributions}\label{sec:stable_condition}

A linear stability analysis provides the condition that the probability
distribution of $\bm{z}$ eventually converges to a steady state distribution.
This happens only when all the real parts of the eigenvalues of the matrix
$\mathsf{A}$ are positive.
The eigenvalues are given by the solutions of the two quadratic equations of
$\psi$
\begin{equation}
\begin{aligned}
m \psi^2 - (\gamma + i B) \psi + (k + i \kappa) & = 0, \\
m \psi^2 - (\gamma - i B) \psi + (k - i \kappa) & = 0.
\end{aligned}
\end{equation}
By solving the equations, we get the four eigenvalues as
\begin{equation}\label{eq:eigenvalues}
\begin{aligned}
    \psi_1 = \frac{\gamma + i B + \sqrt{\Psi}}{2m}, ~~~
    \psi_2 = \frac{\gamma + i B - \sqrt{\Psi}}{2m},
\end{aligned}
\end{equation}
and the complex conjugates $\psi_1^*$ and $\psi_2^*$,
where $\Psi = (\gamma + i B)^2 - 4m(k + i\kappa)$.
The real parts of the eigenvalues take on the two values
\begin{equation}
    \frac{1}{2m}
    \left( \gamma \pm \sqrt{\frac{|\Psi| + {\rm Re}\{\Psi\}}{2}} \right),
\end{equation}
which should both be positive.
Consequently, the system is stable only when the condition
\begin{equation}\label{eq:AStabCondition}
    \gamma k + \kappa B - \kappa^2 m / \gamma
    = \gamma K > 0
\end{equation}
holds~\cite{Lee:2019wl}.

The steady state distribution of the multivariate Ornstein-Uhlenbeck process
is a multivariate Gaussian with zero mean.
The covariance matrix
$\mathsf{C} = \langle \bm{z}\bm{z}^{\rm T} \rangle_{\rm ss}$
is given by solving the matrix equation
$\mathsf{A}\mathsf{C} + \mathsf{C}\mathsf{A}^{\rm T}
= 2\mathsf{D}$~\cite{Gardiner:2010tp}.
In the end we get the steady state distribution as
\begin{equation}\label{eq:Pss}
    P_{\rm ss}(\bm{z}) = \frac{1}{\sqrt{\det(2\pi\mathsf{C})}} 
    \exp \left( -\frac{1}{2}\bm{z} \cdot \mathsf{C}^{-1}\bm{z} \right),
\end{equation}
where
\begin{equation}\label{eq:matC}
    \mathsf{C}
    = \frac{T}{\gamma K} \begin{pmatrix}
    \gamma & 0 & 0 & -\kappa \\
    0 & \gamma & \kappa & 0 \\
    0 & \kappa & (\gamma k + \kappa B)/m & 0 \\
    -\kappa & 0 & 0 & (\gamma k + \kappa B)/m
    \end{pmatrix}.
\end{equation}

In polar coordinates, the position of the particle is identified
by the radial distance from the origin $r = |\bm{x}|$ and the polar angle 
$\theta = \tan^{-1}(x_2/x_1)$.
By changing variables and then marginalizing them, we get the 
probability distribution of the radial distance $r$ as
\begin{equation}\label{eq:radial_dist}
    P_{\rm ss}(r)
    = \frac{K r}{T}
    \exp\left( - \frac{K r^2}{2 T} \right)
\end{equation}
and the conditional distribution of the angular velocity $\dot{\theta}$ for 
a given $r$ as
\begin{equation}\label{eq:theta_dot_dist}
    P_{\rm ss}(\dot{\theta}|r)
    = \sqrt{\frac{mr^2}{2\pi T}} \exp \left[
    - \frac{m r^2}{2 T} \left( \dot{\theta} + \frac{\kappa}{\gamma} \right)^2
    \right].
\end{equation}
From \eqref{eq:radial_dist} and \eqref{eq:theta_dot_dist}, we arrive at
\begin{equation}
    \langle \dot{W} \rangle_{\rm ss}
    = -\kappa \langle r^2 \dot{\theta} \rangle_{\rm ss}
    = \frac{\kappa^2}{\gamma} \langle r^2 \rangle_{\rm ss}
    = \frac{2\kappa^2 T}{\gamma K}
\end{equation}
with $K$ as defined in equation \eqref{eq:AStabCondition}.

%%%%%%%%%%%%%%%%%%%%%%%%%%%%%%%%%%%%%%%%%%%%%%%%%%%%%%%%%%%%%%%%%%%%%%%%%%%%%
\section{Exact expressions of mean value and diffusion coefficient}\label{sec:exact_expression}
The exact mean value and diffusion coefficient of the work $W(t)$ can be obtained
from the scaled cumulant generating function
\begin{equation}\label{eq:SCGF}
    \lambda(h) = \lim_{t\to\infty} \frac{1}{t}
    \ln \left\langle e^{h W(t)} \right\rangle_{\rm ss}.
\end{equation}
For brevity, we rewrite the rate of work as
$\dot{W}(t) = \dot{\bm{z}}(t) \circ \mathsf{W}\bm{z}(t)$ with
\begin{equation}
    \mathsf{W} = \begin{pmatrix}
    0 & \kappa & 0 & 0 \\
    -\kappa & 0 & 0 & 0 \\
    0 & 0 & 0 & 0 \\
    0 & 0 & 0 & 0
    \end{pmatrix}.
\end{equation}
The scaled cumulant generating function is equivalent to the largest 
eigenvalue of the tilted Fokker-Planck 
operator~\cite{Touchette:2018fb}
\begin{equation}
    \mathcal{L}^\dagger
    = -\bm{z} \cdot \mathsf{A}^{\rm T} 
    \left( \bm{\nabla}_{\bm{z}} + h\mathsf{W}\bm{z} \right)
    + \left( \bm{\nabla}_{\bm{z}} + h\mathsf{W}\bm{z} \right)
    \cdot \mathsf{D} \left( \bm{\nabla}_{\bm{z}} + h\mathsf{W}\bm{z}\right).
\end{equation}

We use the Gaussian ansatz
\begin{equation}
	g(\bm{z},h) = \exp \left(-\frac{1}{2}\bm{z} \cdot \mathsf{G}(h)\bm{z}\right)
\end{equation}
for the left eigenfunction of the Fokker-Planck operator 
corresponding to the largest eigenvalue $\lambda(h)$ 
with a symmetric matrix $\mathsf{G}(h)$~\cite{Pietzonka:2018de}.
Since the untilted Fokker-Planck operator preserves probability, the constant function
is the left eigenfunction with largest eigenvalue for $z=0$ and thus $\mathsf{G}(0) = 0$.
From the eigenvalue equation $\mathcal{L}^\dagger g(\bm{z},h) = \lambda(h) g(\bm{z},h)$,
the scaled cumulant generating function follows as
\begin{equation}
    \lambda(h)
    = {\rm tr}\{\mathsf{D}[h\mathsf{W} - \mathsf{G}(h)]\} .
\end{equation}
The matrix $\mathsf{G}(h)$ is determined by the identity
\begin{equation}\label{eq:id_eq}
    \mathsf{A}^{\rm T}(h\mathsf{W} - \mathsf{G})
    + (h\mathsf{W} - \mathsf{G})^{\rm T} \mathsf{A}
    = 2(h\mathsf{W} - \mathsf{G})^{\rm T} \mathsf{D}
    (h\mathsf{W} - \mathsf{G}).
\end{equation}

It is hard to determine the matrix $\mathsf{G}(h)$ by solving the 
nonlinear equation in \eqref{eq:id_eq}.
Nevertheless, the cumulants can be obtained from
a series expansion of $\lambda(h)$ at $h = 0$.
Provided that the matrix $\mathsf{G}(h)$ is expanded as a series 
$\mathsf{G}(h) = \sum_{n=1}^\infty \mathsf{G}_{n} h^n$,
the mean currents and diffusion coefficients are given by 
\begin{equation}
\langle \dot{W} \rangle_{\rm ss}
=\left. \frac{\partial\lambda}{\partial h} \right|_{h=0}
= {\rm tr}\{ \mathsf{D}(\mathsf{W} - \mathsf{G}_{1})\}
\end{equation}
and
\begin{equation}
D_W
= \frac{1}{2} 
\left. \frac{\partial^2\lambda}{\partial h^2} \right|_{h=0}
= -{\rm tr}\{ \mathsf{D}\mathsf{G}_{2}\},
\end{equation}
respectively.
The matrices $\mathsf{G}_{n}$ are determined by solving \eqref{eq:id_eq}
order by order in powers of $h$.
The first two orders give the identities
\begin{equation}
\begin{aligned}
    \mathsf{A}^{\rm T}(\mathsf{W} - \mathsf{G}_1)
    + (\mathsf{W} - \mathsf{G}_1)^{\rm T}\mathsf{A} & = 0, \\
    \mathsf{A}^{\rm T}\mathsf{G}_2 + \mathsf{G}_2\mathsf{A}
    + 2(\mathsf{W} -\mathsf{G}_1)^{\rm T}\mathsf{D}
    (\mathsf{W} -\mathsf{G}_1) & = 0.
\end{aligned}
\end{equation}
From these identities and the relation $\mathsf{A}\mathsf{C}
+ \mathsf{C}\mathsf{A}^{\rm T} = 2\mathsf{D}$, we get
\begin{equation}
    \langle \dot{W} \rangle_{\rm ss}
    = {\rm tr} \{\mathsf{A}\mathsf{C} \mathsf{W}_a \},
\end{equation}
and
\begin{equation}
    D_W = {\rm tr}\{ \mathsf{A} \mathsf{C} \mathsf{W}_a \mathsf{C}
    (\mathsf{W}^{\rm T} - \mathsf{G}_1) \},
\end{equation}
where $\mathsf{W}_a = (\mathsf{W}-\mathsf{W}^{\rm T})/2$ is the asymmetric
part of the matrix $\mathsf{W}$.
Thus we get the exact expressions of the mean current and the diffusion 
coefficient of the work current by finding $\mathsf{G}_1$, which are given
in \eqref{eq:wss} and \eqref{eq:Dw} in terms of the specified 
dimensionless parameters.

Analogously the heat current can also be rewritten
as $\dot{Q}(t) = \dot{\bm{z}}(t)\circ \mathsf{Q}\bm{z}(t)$ with
\begin{equation}
    \mathsf{Q} = \begin{pmatrix}
    -k & \kappa & 0 & B \\
    -\kappa & -k & -B & 0 \\
    0 & 0 & -m & 0 \\
    0 & 0 & 0 & -m
    \end{pmatrix}.
\end{equation}
The mean value and diffusion coefficient of the heat current are given by
a similar procedure in which the matrix $\mathsf{W}$ is replaced with 
$\mathsf{Q}$.
The mean value and diffusion coefficient of the heat current are same
with those of the work current.
Consequently, the uncertainty product of the heat current is same as 
that of the work current.

%%%%%%%%%%%%%%%%%%%%%%%%%%%%%%%%%%%%%%%%%%%%%%%%%%%%%%%%%%%%%%%%%%%%%%%%%%%%%
\section{Scaling behavior of diffusion coefficient of work current}\label{sec:scaling}

In the following, we analyze the scaling behavior of the diffusion coefficient
of the work current. We do this by rewriting the diffusion coefficient \eqref{eq:Dw_def}
as a Green-Kubo expression
\begin{equation}\label{eq:Dw_int}
    D_W
    = \lim_{t\to\infty} \int_0^t ds
    \langle \delta \dot{W}(s) \delta \dot{W}(t) \rangle_{\rm ss},
\end{equation}
where $\delta \dot{W}(t) = \dot{W}(t) - \langle \dot{W}(t) \rangle$.
The correlation function of the work current
$\langle \delta\dot{W}(s) \delta\dot{W}(t) \rangle_{\rm ss}$
can be expressed as a sum of four-point correlation functions of $\bm{z}(t)$.
Since the steady state distribution and transition probability 
of $\bm{z}(t)$ are Gaussian, the four-point correlation 
functions can be divided into two-point correlation functions using
Wick's theorem.
Thus we have
\begin{equation}\label{eq:work-work_corr}
\begin{aligned}
    \langle \delta\dot{W}(s) \delta\dot{W}(t) \rangle_{\rm ss}
    & =  2\kappa^2 
    ( \langle x_1(s)x_1(t) \rangle_{\rm ss}
    \langle v_1(s)v_1(t) \rangle_{\rm ss} \\
    & ~~~~~~~
    + \langle x_1(s)x_2(t) \rangle_{\rm ss}
    \langle v_1(s)v_2(t) \rangle_{\rm ss} \\
    & ~~~~~~~
    - \langle x_1(s)v_2(t) \rangle_{\rm ss}
    \langle v_1(s)x_2(t) \rangle_{\rm ss} \\
    & ~~~~~~~
    - \langle x_1(s)v_1(t) \rangle_{\rm ss}
    \langle v_1(s)x_1(t) \rangle_{\rm ss} 
    )
\end{aligned}
\end{equation}
where the relations $\langle x_1(s) x_1(t) \rangle_{\rm ss}
= \langle x_2(s) x_2(t) \rangle_{\rm ss}$,
$\langle x_2(s)x_1(t) \rangle_{\rm ss}
= -\langle x_1(s)x_2(t) \rangle_{\rm ss}$, and their time derivatives
are used.
These relations come from the fact that the physics of the system
is not affected by the relabelling coordinates $(x_1,x_2) \to (x_2,-x_1)$
due to the spatial isotropy of the system.
Additionally, an integration by parts leads to
\begin{equation}\label{eq:Dw_appendix}
\begin{aligned}
    D_W & = 4\kappa^2 \lim_{t\to\infty} \int_0^t ds
    (\langle x_1(s)x_1(t) \rangle_{\rm ss} \langle v_1(s) v_1(t)
    \rangle_{\rm ss} \\
    & ~~~~~~~~~~~~~~~~~~~
    + \langle x_1(s) x_2(t) \rangle_{\rm ss} \langle v_1(s)v_2(t)
    \rangle_{\rm ss} ).
\end{aligned}
\end{equation}

The two-point correlation functions of $\bm{z}(t)$ involving two different 
times $s$ and $t$ are determined by the matrices $\mathsf{A}$ and $\mathsf{C}$ as
$\langle \bm{z}(t-\tau) \bm{z}^{\rm T}(t) \rangle_{\rm ss}
= \langle \bm{z}(t) \bm{z}^{\rm T}(t+\tau) \rangle_{\rm ss}
= \mathsf{C} e^{-\mathsf{A}^{\rm T}\tau}$.
The matrix exponential can be decomposed as
\begin{equation}\label{eq:exp_decomp}
    e^{-\mathsf{A}^{\rm T} \tau}
    = \sum_{j=1}^2 \left(
    e^{\psi_j\tau}\bm{l}_j\bm{r}_j^{\rm T}
    + e^{\psi_j^* \tau}\bm{l}_j^* (\bm{r}_j^*)^{\rm T}\right)
\end{equation}
where $\bm{l}_{j}$ and $\bm{r}_j$ are the left and
right eigenvector corresponding to the eigenvalue $\psi_j$, respectively.
The eigenvectors are orthonormal,
namely, $\bm{l}_j\cdot\bm{r}_k = \delta_{jk}$ and
$\bm{l}_j \cdot \bm{r}_k^* = 0$.
Up to normalization, they are given by
\begin{equation}
\begin{aligned}
    \bm{l}_1 &
    \propto (\psi_2, i\psi_2, 1, i)^{\rm T}, \\
    \bm{l}_2 &
    \propto (\psi_1, i\psi_1, 1, i)^{\rm T}, \\
    \bm{r}_1 & \propto (1, -i, -\psi_1, i\psi_1)^{\rm T}, \\
    \bm{r}_2 & \propto (1, -i, -\psi_2, i\psi_2)^{\rm T}.
\end{aligned}
\end{equation}
Apart from those eigenvectors the correlation function has time dependent
contributions with different relaxation times and oscillation frequencies,
which are determined by the eigenvalues $\psi_1$ and $\psi_2$.
Thus the time dependence of
$\langle \delta\dot{W}(t-\tau)\delta\dot{W}(t) \rangle_{\rm ss}$
is determined by $e^{-(\psi_j + \psi_k)\tau}$ and
$e^{-(\psi_j + \psi^*_k)\tau}$ for all pairs of $(j,k)$
with some prefactors.

In the strong magnetic field limit, the eigenvalues of $\mathsf{A}$
up to the leading order are given by $\psi_1 = (\gamma + i B)/m$,
$\psi_2 = (\kappa - i k)/B$ and their complex conjugates.
The two-point correlation function of
the position and that of the velocity up to the order of $B^{-2}$ are given by
\begin{widetext}
\begin{equation}\label{eq:xx_corr}
    \langle \bm{x}(t-\tau)\bm{x}^{\rm T}(t) \rangle_{\rm ss}
    \simeq \frac{m T}{B^2} e^{-\gamma \tau/m}
    \mathsf{R}\left( B \tau / m \right)
    + \frac{\gamma T}{\kappa B}
    \left( 1 + \frac{\gamma k}{\kappa B} \right) e^{-\kappa\tau/B}
    \mathsf{R}\left( k \tau / B \right),
\end{equation}
\begin{equation}\label{eq:vv_corr}
    \langle \bm{v}(t-\tau)\bm{v}^{\rm T}(t) \rangle_{\rm ss}
    \simeq T \left( 
    \frac{1}{m}
    + \frac{\gamma k + \kappa B - \kappa^2 m /\gamma}{\gamma B^2} \right)
    e^{-\gamma \tau /m} \mathsf{R}(B \tau/m)
\end{equation}
\end{widetext}
with
\begin{equation}
    \mathsf{R}(\varphi) = \begin{pmatrix}
        \cos\varphi & -\sin\varphi \\
        \sin\varphi & \cos\varphi
    \end{pmatrix}.
\end{equation}

The leading terms of the cross correlation functions
$\langle \bm{x}(t-\tau)\bm{v}^{\rm T}(t) \rangle_{\rm ss}$ and 
$\langle \bm{v}(t-\tau)\bm{x}^{\rm T}(t) \rangle_{\rm ss}$ are 
of the order $B^{-1}$.
Thus the correlation function of the work current in \eqref{eq:work-work_corr}
is primarily determined by
\begin{align}\label{eq:BleadingORder}
	\langle x_1(t- & \tau)x_1(t) \rangle_{\rm ss}
	\langle v_1(t-\tau)v_1(t) \rangle_{\rm ss} \nonumber \\
	& \simeq (\gamma T^2/(\kappa m B))
	e^{-\gamma \tau/m}\cos(B\tau/m)	
\end{align}
which like the velocity and position correlation functions follows oscillations
that are decaying over time. The integration of this leading term over $\tau$ gives
a result of the order $B^{-3}$. This is, however, a subleading term of the diffusion coefficient.

The leading term of the diffusion coefficient arises from the integration
of subleading, additional terms that correspond to $e^{-(\psi_1+\psi_1^*)\tau}$.
The product of the first term of \eqref{eq:xx_corr} and the leading term
of \eqref{eq:vv_corr} gives
\begin{equation} \label{eq:Bintegration1}
\begin{aligned}
    I_1 & = \int_0^\infty d\tau
    \langle x_1(t-\tau)x_1(t)\rangle_{\rm ss}
    \langle v_1(t-\tau)v_1(t)\rangle_{\rm ss} \\
    & = \frac{T^2}{B^2} \int_0^\infty d\tau
    e^{-2\gamma \tau/m} \cos^2\left(\frac{B\tau}{m}\right)
    + \mathcal{O}(B^{-3}),
\end{aligned}
\end{equation}
\begin{equation} \label{eq:Bintegration2}
\begin{aligned}
    I_2 & = \int_0^\infty d\tau
    \langle x_1(t-\tau)x_2(t)\rangle_{\rm ss}
    \langle v_1(t-\tau)v_2(t)\rangle_{\rm ss} \\
    & = \frac{T^2}{B^2} \int_0^\infty d\tau
    e^{-2\gamma \tau/m} \sin^2\left(\frac{B\tau}{m}\right)
    + \mathcal{O}(B^{-3}),
\end{aligned}
\end{equation}
and finally
\begin{equation}
\begin{aligned}
    D_W = 4\kappa^2(I_1 + I_2)
    = \frac{2\kappa^2 m T^2}{\gamma B^2} + \mathcal{O}(B^{-3}).
\end{aligned}
\end{equation}
The oscillations do not cancel in the integrations \eqref{eq:Bintegration1} and
\eqref{eq:Bintegration2} since the integrands are positive.
The mean work current scales as $\langle \dot{W} \rangle_{\rm ss} \sim B^{-1}$
for large $B$ provided that $\kappa B >0$.
Thus the squared relative uncertainty
$\epsilon^2_W = 2 D_W/\langle \dot{W} \rangle^2_{\rm ss}$ converges to
a constant as $B$ increases.
In the small mass limit $mB \ll 1$, the leading order of the diffusion 
coefficient becomes $B^{-3}$, which results in a decreasing uncertainty
$\epsilon^2_W \sim B^{-1}$.

\bibliographystyle{apsrev}
\bibliography{paper}

\end{document}